\documentclass[9pt,twocolumn,twoside]{osajnl}

\journal{ol} 

\setboolean{shortarticle}{true}

\title{Real-Time Optical Time Interpolation Using Spectral Interferometry}

\author[1,*]{Thomas Fordell}

\affil[1]{VTT Technical Research Centre of Finland Ltd, National Metrology Institute VTT MIKES, Tekniikantie 1, 02150, Finland}

\affil[*]{Corresponding author: thomas.fordell@vtt.fi}

 \doi{\url{https://doi.org/10.1364/OL.450266}}

\begin{abstract}
A simple scheme for all-optical time interpolation using spectral interferometry is put forward that is in principle capable of single-shot measurements. In this method, the arrival time of optical timing pulses is encoded into the spectrum of a time-stretched supercontinuum via cross-phase modulation. The proof-of-concept test setup points toward femtosecond-level absolute timing capabilities with only minor additions to modern optical clockwork.\\

© 2022 Optica Publishing Group. Users may use, reuse, and build upon the article, or use the article for text or data mining, so long as such uses are for non-commercial purposes and appropriate attribution is maintained. All other rights are reserved.
\end{abstract}

\begin{document}

\maketitle

\section{Introduction}
Time interval counters (or event timers) and frequency counters are vital instruments in several fields of science, including time and frequency metrology, geodesy, astronomy, as well as materials, nuclear and particle physics. 
High-resolution instruments rely on time interpolation, which enables the triggering time to be interpolated in between cycles of the reference clock, thereby increasing the resolution by several orders of magnitude. Many different schemes for realizing electrical time interpolation have been put forward over the years \cite{Kalisz2003a}
, and with careful design, a single-shot precision of about 1~ps is possible (e.g. \cite{Keranen2011a}).

\begin{figure}[t]
\centering
\includegraphics[width=0.8\linewidth]{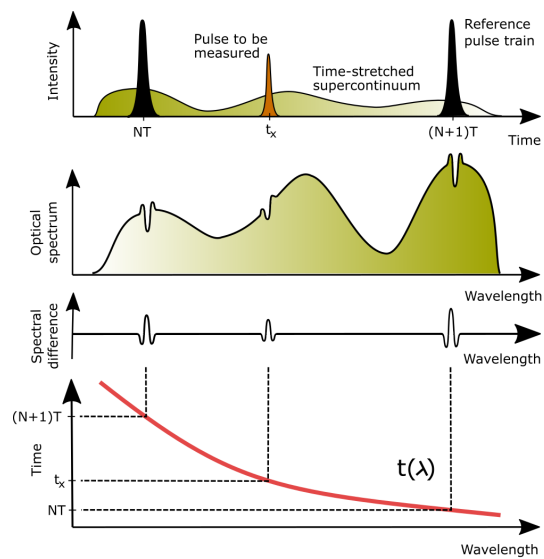}\\
\caption{Illustration of optical time interpolation in between  time marker pulses from a mode-locked laser. The moment of time is encoded into the spectrum of a time-stretched supercontinuum via cross-phase modulation.}
\label{fig:schematic}
\end{figure}

For the last two decades, time and frequency metrology has been moving from the electrical domain to the optical domain: 
optical frequency standards have been realized with fractional frequency uncertainty estimates at the 10$^{-18}$ level 
 and  optical clockwork (frequency combs) enable frequency measurements at the 20th decimal place.  
With these developments, all optical time-interpolation becomes a relevant issue. 

Dual-comb asynchronous optical sampling \cite{Schiller2002a} has turned out to be a very powerful spectroscopic technique where a pulse train with a slightly detuned repetition rate ($f_{r} + \Delta f$) is scanned across the signal pulses ($f_r$) on a photo-detector and processed. This effectively stretches the time between the pulses ($1/f_r$) to the measurement time ($1/\Delta f$) with an effective sampling time of $\Delta t=\Delta f/(f_r (f_r + \Delta f)$. This allows for extremely high temporal resolution provided sufficient coherence can be maintained between the pulse trains. Recently, dual-comb techniques have been used for high-precision time transfer demonstrations over a 4-km free-space link \cite{Deschenes2016a} and a 114-km optical fibre link \cite{Abuduweili2020a}.


An elegant real-time  approach for high-resolution measurements of optical waveforms is temporal imaging (magnification) using a time lens (e.g. \cite{Kolner1989a, Foster2008a, Li2021a}). A somewhat related approach is photonic time stretching, which enables high-resolution measurements of electrical signals \cite{Chou2007a}. According to \cite{Galvanauskas1992a}, the first attempt to make a real-time optical measurement of an ultrafast process was made in \cite{Valdmanis1986a}, where pulses from a 70-fs mode-locked dye laser were stretched in an optical fibre to 300~ps and passed through an electro-optical modulator placed between two crossed polarizers. Consequently, the time-dependent electrical signal was encoded into the optical spectrum.  
In \cite{Beddard1992a} this approach was applied to transient absorption and in \cite{Jiang1998b} to travelling THz waves. Introduction of a reference pulse either temporally or spatially separated from the probe pulse \cite{Chien2000a, Leblanc2000a, Geindre2001a, Kim2002a, Sharma2012a} enables under certain restrictions \cite{Wahlstrand2016a} using Fourier-transform-based fringe pattern analysis 
the temporal reconstruction of the phase perturbation. In the present work, the objective is to cover a much longer time span (nanoseconds), which means that spectral fringes cannot be resolved and the temporal phase perturbation cannot be reconstructed. Nevertheless, the effective temporal position of the perturbation can be determined with high precision as shown below also without the use of a separate supercontinuum pulse for phase referencing.

The method works as follows. A supercontinuum is dispersively stretched over the time interval to be measured and combined, in, e.g., a nonlinear waveguide, with short optical timing reference pulses and the pulse or pulses to be measured.  Fig.~\ref{fig:schematic} depicts a single period of the continuous process with a supercontinuum that stretches across the pulse period. The timing pulses have to have high enough peak power and/or the interaction length has to be long enough so that measurable nonlinear phase shifts to the supercontinuum occur via  cross-phase modulation (XPM). As will be shown below, XPM will produce fringes that have an easily identifiable 'centre', the wavelength of which can be converted to time via the (calibrated) chirp of the supercontinuum.

\section{Theoretical aspects}

To gain insight, let the complex field of the time-stretched supercontinuum (sc) be $E_{nl}(t)=A_{sc}(t)e^{i\phi_{sc}(t) + i\phi_{nl}(t)}$, where
$\phi_{nl}(t)$ is the nonlinear phase acquired due to XPM by a timing pulse $E_p(t)$. If $\phi_{nl}(t)\ll1$, then $E_{nl}(t)\approx A_{sc}(t)e^{i\phi_{sc}(t)}
(1 + i\phi_{nl}(t))$. If the interaction is assumed  short (no walk-off) and the reference timing pulse is assumed Gaussian $E_p(t)=A_{p}e^{-\alpha_p^2 (t-t_0)^2}$, then the spectrum becomes
\begin{eqnarray}
|E_{nl}(\omega)|^2 &\approx& |E_{sc}(\omega)|^2 + \eta E_{sc}(\omega)  e^{-\frac{\Delta\omega^2}{4\alpha_p^2}-i\pi/2} + c.c.,
\end{eqnarray}
where $\eta=A_{sc}(t_0)\phi^{max}_{nl}\sqrt{\pi}\alpha_p^{-1}$.
After subtracting the supercontinuum spectrum (without XPM),  dividing by its square root and neglecting a contribution from the chirp of the supercontinuum, which for the present experimental parameters is tiny, the signal $S$ becomes
\begin{equation} 
S = \frac{|E_{nl}(\omega)|^2-|E_{sc}(\omega)|^2}{\sqrt{|E_{sc}(\omega)|^2}}=2\eta \textrm{cos}(\frac{\beta_2 L}{2} \Delta\omega^2 + \frac{\pi}{4})e^{-\frac{\Delta\omega^2}{4\alpha_p^2}},
\label{Eq:signal}
\end{equation}
where $\beta_2 L$ is the group-delay dispersion. Thus, the width of the modulation is set by the bandwidth of the timing pulse (assumed smaller than the supercontinuum) and the modulation phase increases proportional to $\Delta \omega^2$. In practice, the range of visibility of the fringes will be limited by the resolution of the spectrometer. 

\begin{figure}
\includegraphics[width=0.5\linewidth]{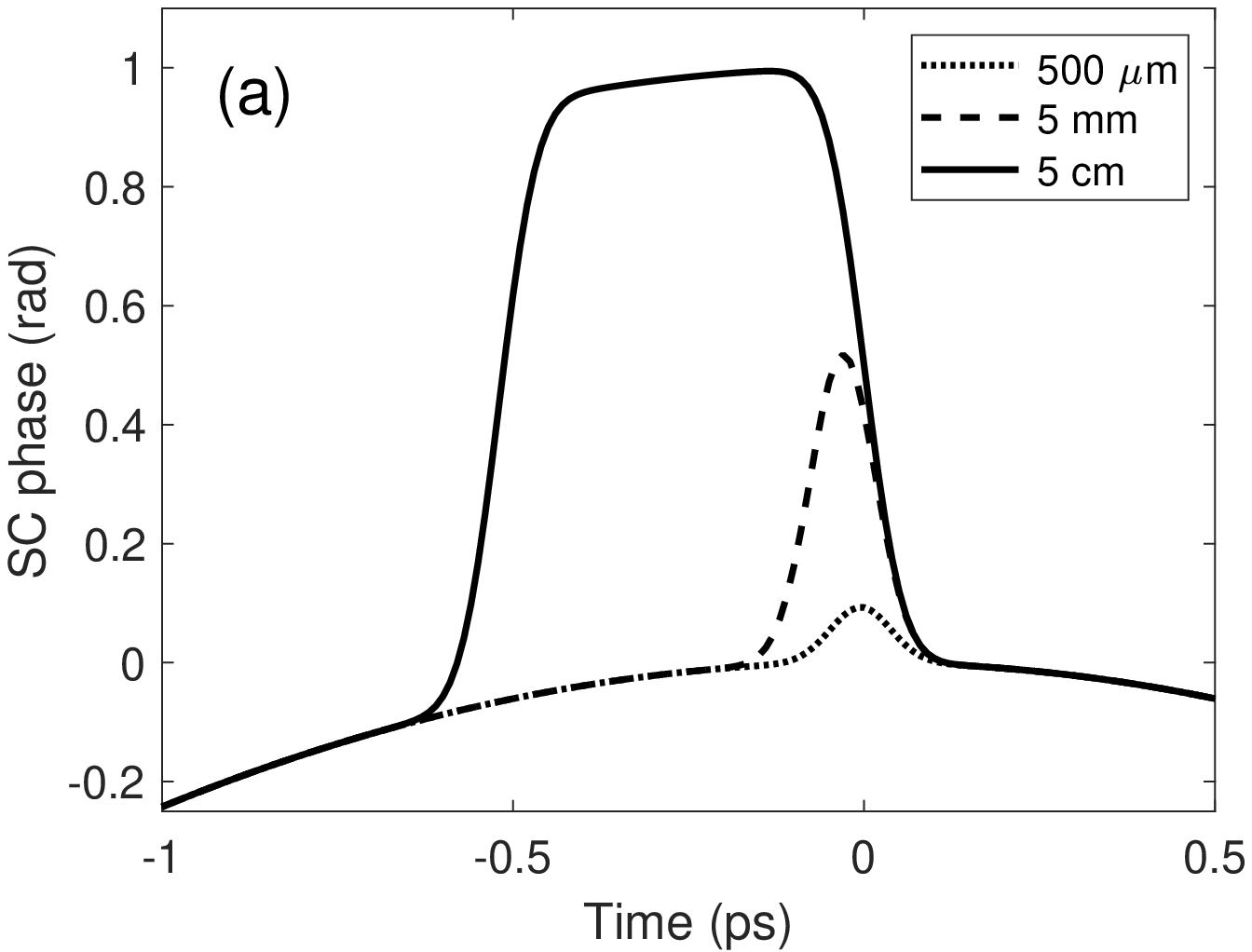}
\includegraphics[width=0.5\linewidth]{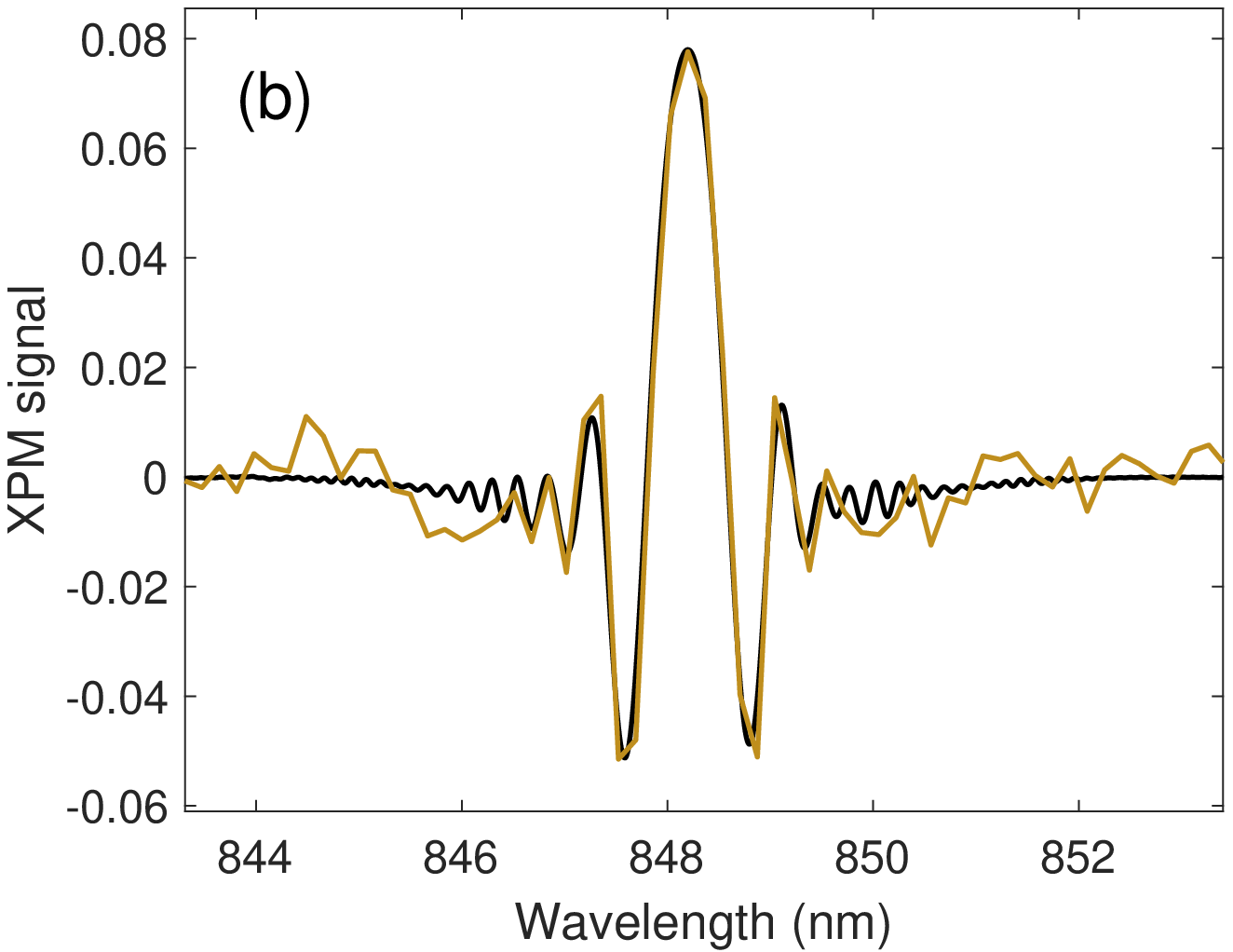}
\caption{(a) Nonlinear phase shifts induced by cross-phase modulation on the chirped supercontinuum for different interaction lengths (b) Example of experimental (brown) and numerical (black) XPM signal.}
\label{fig:nonlinear_phase}
\end{figure}
The resulting temporal phase of the scenario above is depicted by the dotted line in Fig.~\ref{fig:nonlinear_phase}a, which shows a small Gaussian phase shift at $t=0$ on top of the quadratic phase profile of the chirped supercontinuum; however, in practice it is desirable to have $\phi^{max}_{nl}\approx1$, and if the interaction takes place over an extended length as in this work (solid line), considerable 'stretching' of the phase shift will take place due to group velocity mismatch (pulse walk-off). Numerical integration \cite{AgrawalBook} with parameters corresponding to the setup described in the next section, results in the XPM signal shown in Fig.~\ref{fig:nonlinear_phase}b (black line), which is in good agreement with the experimental trace with only the vertical scaling as a free parameter. Here the fringe visibility is  limited by the 0.2-nm resolution of the spectrometer.

\section{Experiment}

\begin{figure}[t]
\includegraphics[width=\linewidth]{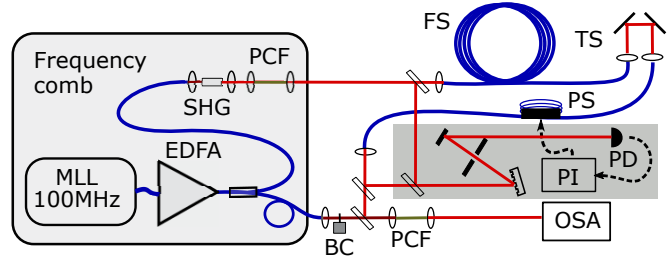}
\caption{Experimental setup. MLL: mode-locked laser; SHG: second-harmonic generation; PCF: photonic crystal fiber; BC: beam chopper; FS: PM-fibre spool; TS: translation stage; PS: piezo stretcher; PI:proportional-integral controller; PD: photodiode; OSA: optical spectrum analyzer.}
\label{fig:setup}
\end{figure}

The test setup is shown in Fig. \ref{fig:setup}. 
The near-infrared supercontinuum from a frequency doubled 100-MHz frequency comb is time-stretched in a 54-m polarization-maintaining fibre (PM780) and focused into a 5-cm long PCF (PM-1550, NKT Photonics) along with the fundamental pulse train at 1560~nm picked off after the amplifier. A compact, low-cost  spectrometer (Thorlabs CCS175) with a $\sim 0.2$-nm resolution (at 850~nm) was used to record spectra. A translation stage (TS) was used to vary the temporal overlap between the pulses and a beam chopper (BC) enabled the recording of reference spectra.

\begin{figure}[t]
\includegraphics[width=\linewidth]{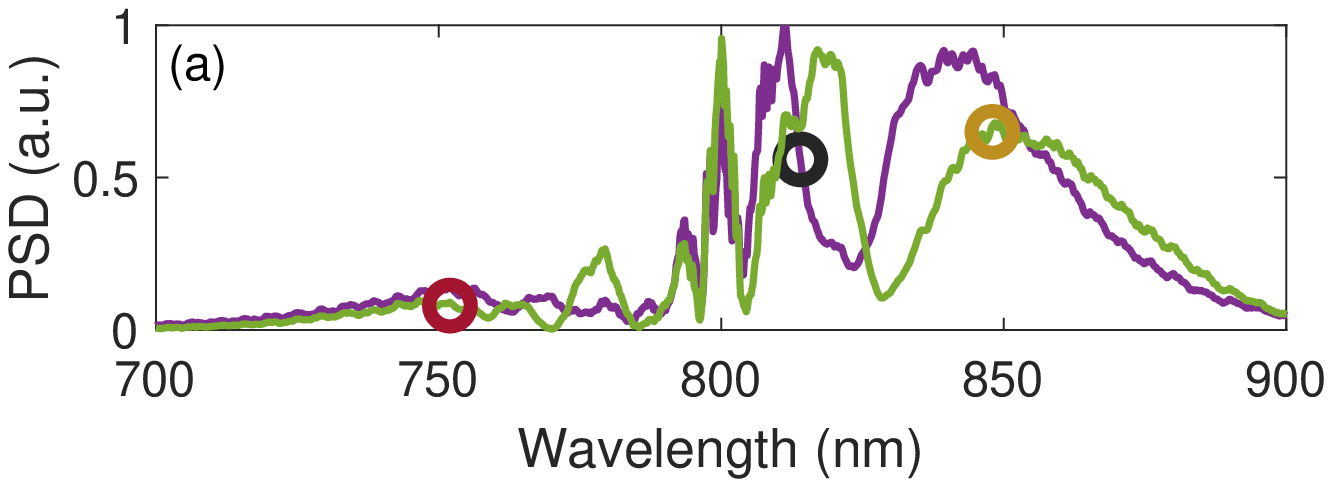}
\includegraphics[width=\linewidth]{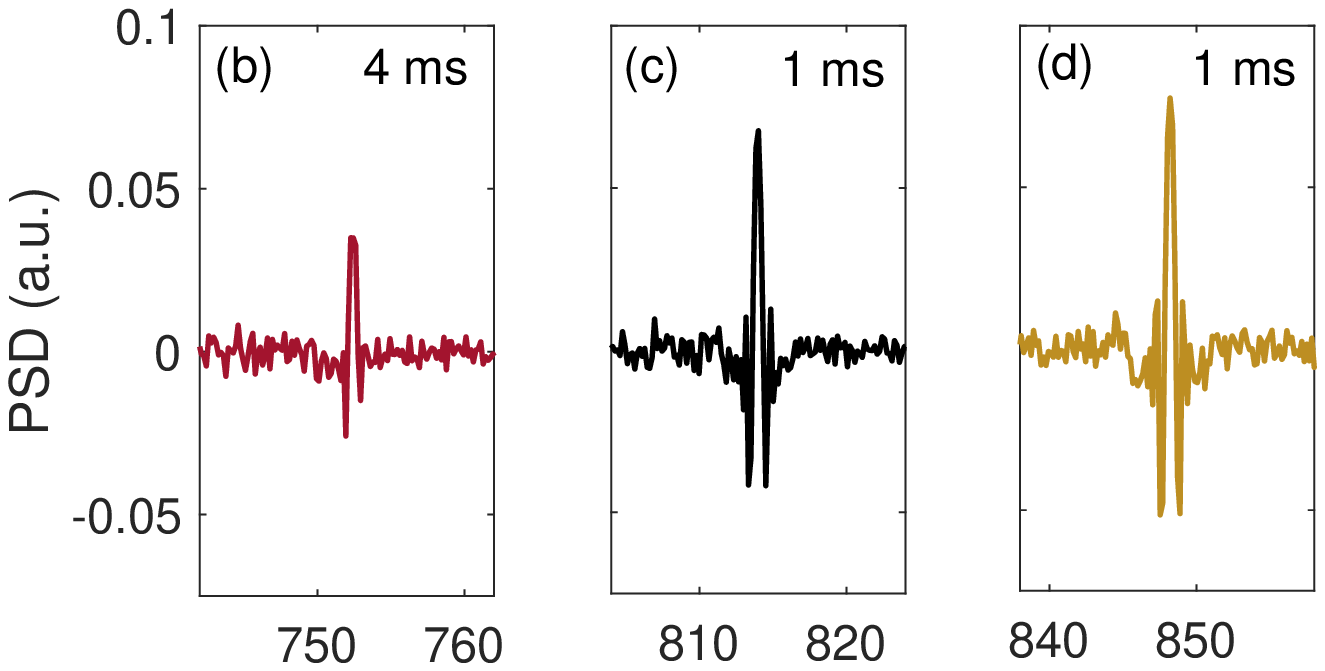}
\caption{Example spectra of the supercontinua used in this work (top) and the corresponding XPM signals (bottom) obtained at the positions marked with circles. The OSA integration time is also given.}
\label{fig:spectra}
\end{figure}
The laser is mode-locked by nonlinear polarization rotation, and fiber squeezers are used to adjust the mode-locked state and also the amount of light in the two branches. This, along with drifts in the free-space sections causes the supercontinuum spectrum to drift from day to day. Fig. \ref{fig:spectra}a shows two reference spectra recorded on different days, and Figs. \ref{fig:spectra}b-d show the XPM signal corresponding to the colored markers in Fig.\ref{fig:spectra}a. The spectrometer saturated at integration times longer than about 1~ms. Nevertheless, the XPM signal at 752~nm was recorded with a 4-ms integration time in order to make it more clear. These traces were obtained by first recording a reference spectrum and then immediately afterwards the XPM signal spectrum, subtracting these two and finally also subtracting a low-pass filtered copy. This last step was needed since the chopper, while being 'low-noise', still caused some mechanical disturbance that shifted the spectra slightly, causing a modulation on the trace. Finally, the position of the peaks was determined by a 5-point Gaussian fit. 

\begin{figure}[t]
\centering
\includegraphics[width=\linewidth]{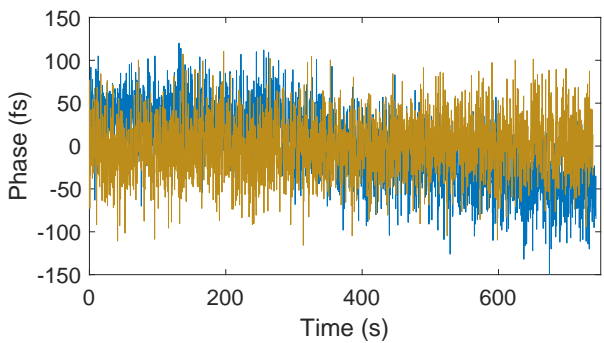}\\
\caption{Measured timing phase at 848~nm with (brown) and without active stabilization of the fiber delay line.}
\label{fig:phase_time_series}
\end{figure}

During longer measurement times, a clear drift in the peak positions could be seen (blue trace in Fig.~\ref{fig:phase_time_series}). An obvious culprit is the long delay line. To verify this, a small part of the supercontinuum was split off before and after the delay line. After spectral filtering, interference was observed on a photodetector allowing a PI-controller to stabilize the delay via a fiber stretcher (gray box in Fig. \ref{fig:setup}). The yellow trace in Fig.~\ref{fig:phase_time_series} shows a clear reduction in the drift when feedback was on. It should be noted that in the current setup this still leaves several meters of unstabilized beam paths inside the frequency comb.   
  
\section{Results}

\begin{figure}[t]
\centering
\includegraphics[width=0.9\linewidth]{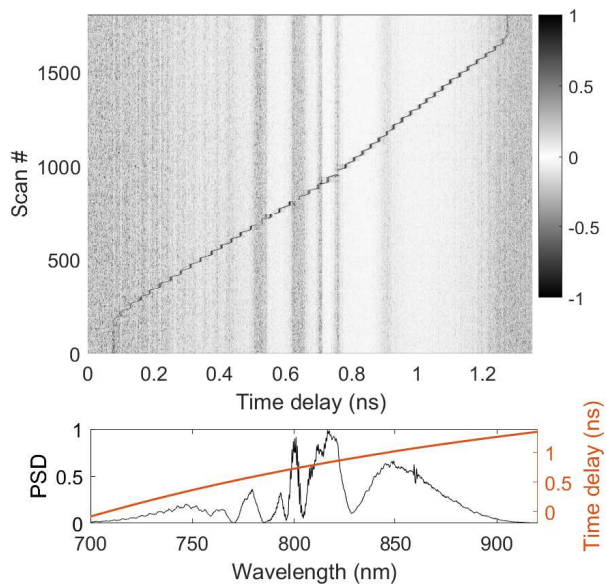}\\
\caption{Pseudo-color plot (top) of the XPM signal as the delay is scanned manually across the supercontinuum. The bottom graph shows the spectrum (black) where the XPM signal is visible at 860~nm. The red line shows the delay dependence.}
\label{fig:scan}
\end{figure}

A false-color plot of the normalized XPM signal as the delay is scanned across the supercontinuum spectrum is shown in Fig.~\ref{fig:scan}a. In this scan, the integration time was kept constant at 500~$\mu$s during the entire scan, which explains the poor signal to noise at the lower wavelengths. Fig.~\ref{fig:scan}b shows a single spectrum (black) with the XPM signal clearly visible at 860~nm. The right y-axis shows the relation between time delay as deduced from the position of the delay stage and the XPM signal peak wavelength.

What is really of interest is the timing of the pulse to be measured with respect to a reference pulse, in which case the supercontinuum is only used for interpolation and the slow drift seen in Fig.~\ref{fig:phase_time_series} (blue trace) becomes irrelevant. In this work such a full system was not yet available, but an indication of the potential performance can be obtained by studying the time instability of a single reference pulse with respect to the supercontinuum.

\begin{figure}[t]
\includegraphics[width=0.5\linewidth]{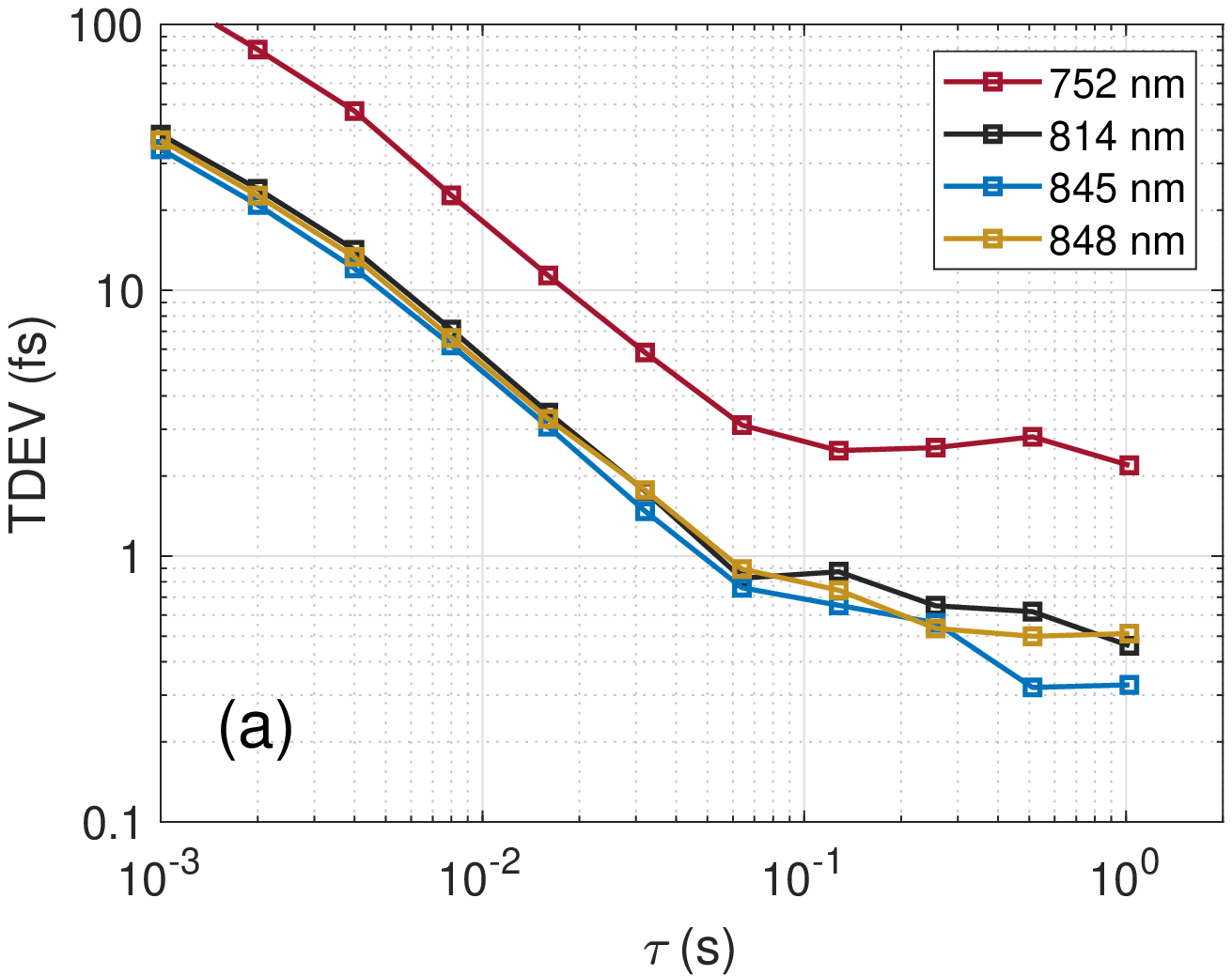}
\includegraphics[width=0.5\linewidth]{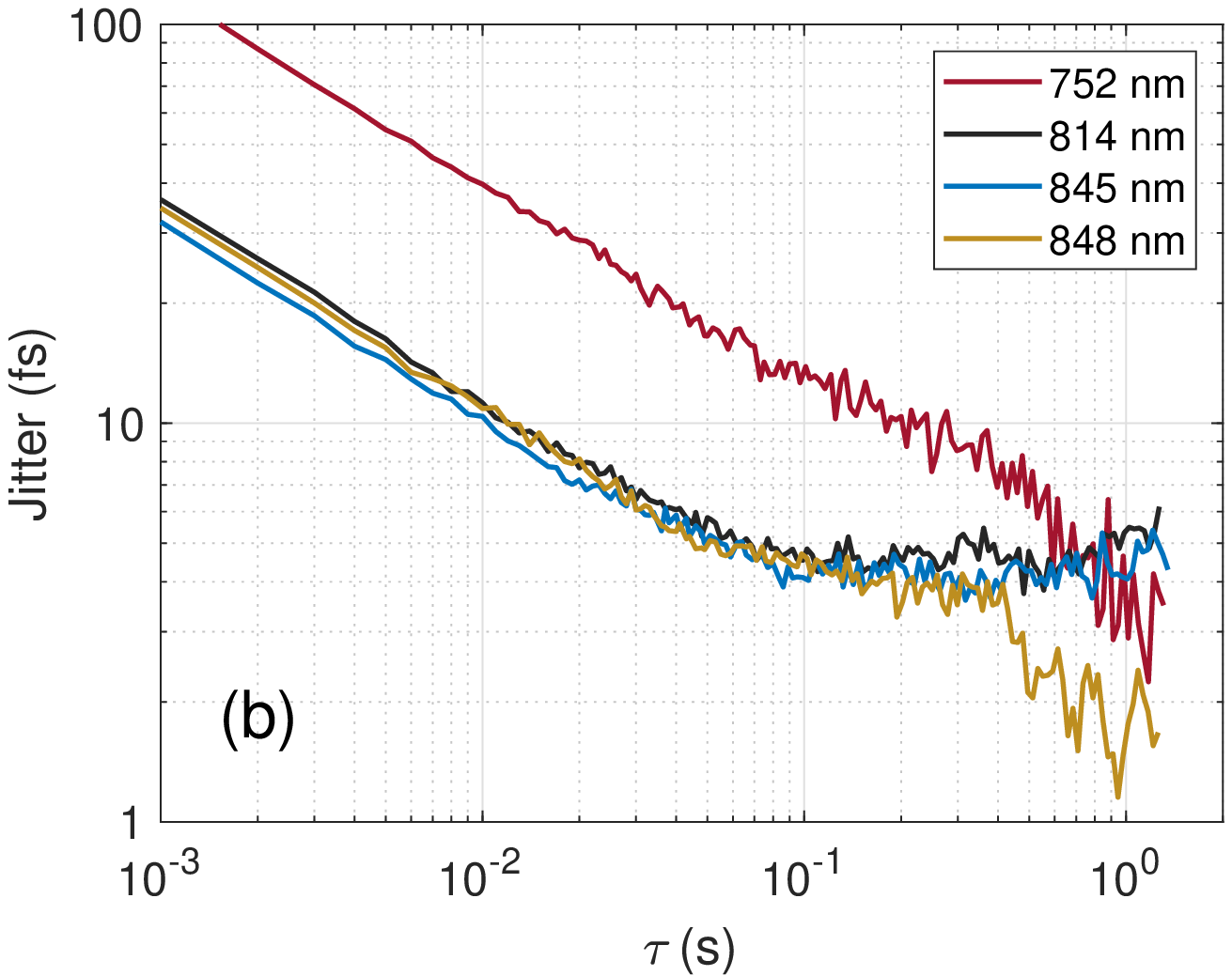}
\caption{(a) Time deviation with dead time. See text for details. (b) Differential jitter as defined in the main text.}
\label{fig:stability}
\end{figure}

Timing instability is usually characterized by the time deviation (TDEV) defined via the modified allan deviation (mod $\sigma_y(\tau))$. The tricky thing here is that $\sigma_y(\tau)$  requires dead-time-free data but the readout rate of the OSA  was only 12~Hz and the integration time was about 1~ms, giving a dead time to integration time ratio of $\sim 80$. While dead time contribution to $\sigma_y(\tau)$ can be taken into account via the $B_2$ and $B_3$ bias functions \cite{Barnes1990a} and while mod~$\sigma_y(\tau)$ can be computed from $\sigma_y(\tau)$ \cite{Sullivan1990a} the results are prone to significant errors due to the sensitive dependence on the noise model particularly for large multiples of the sampling time $\tau_0$. Nevertheless, Fig. \ref{fig:stability}a shows a best estimate for the dead-time-corrected TDEV for three different wavelengths where the program Stable32 was used to obtain the noise models: a white phase noise model was used for small $\tau$ and flicker phase for large $\tau$. To add confidence, Fig. \ref{fig:stability}b shows the jitter defined as the standard deviation of the difference between consecutive phase readings  that have been averaged over time $\tau$ ignoring dead time and that have been normalized by $\sqrt{2}$. At 814~nm and 846~nm, the jitter seems to level off at 4-5~fs; however, when the delay line is actively stabilized (yellow), the jitter so defined keeps falling towards the 1-fs level. 

\begin{figure}[]
\includegraphics[width=0.5\linewidth]{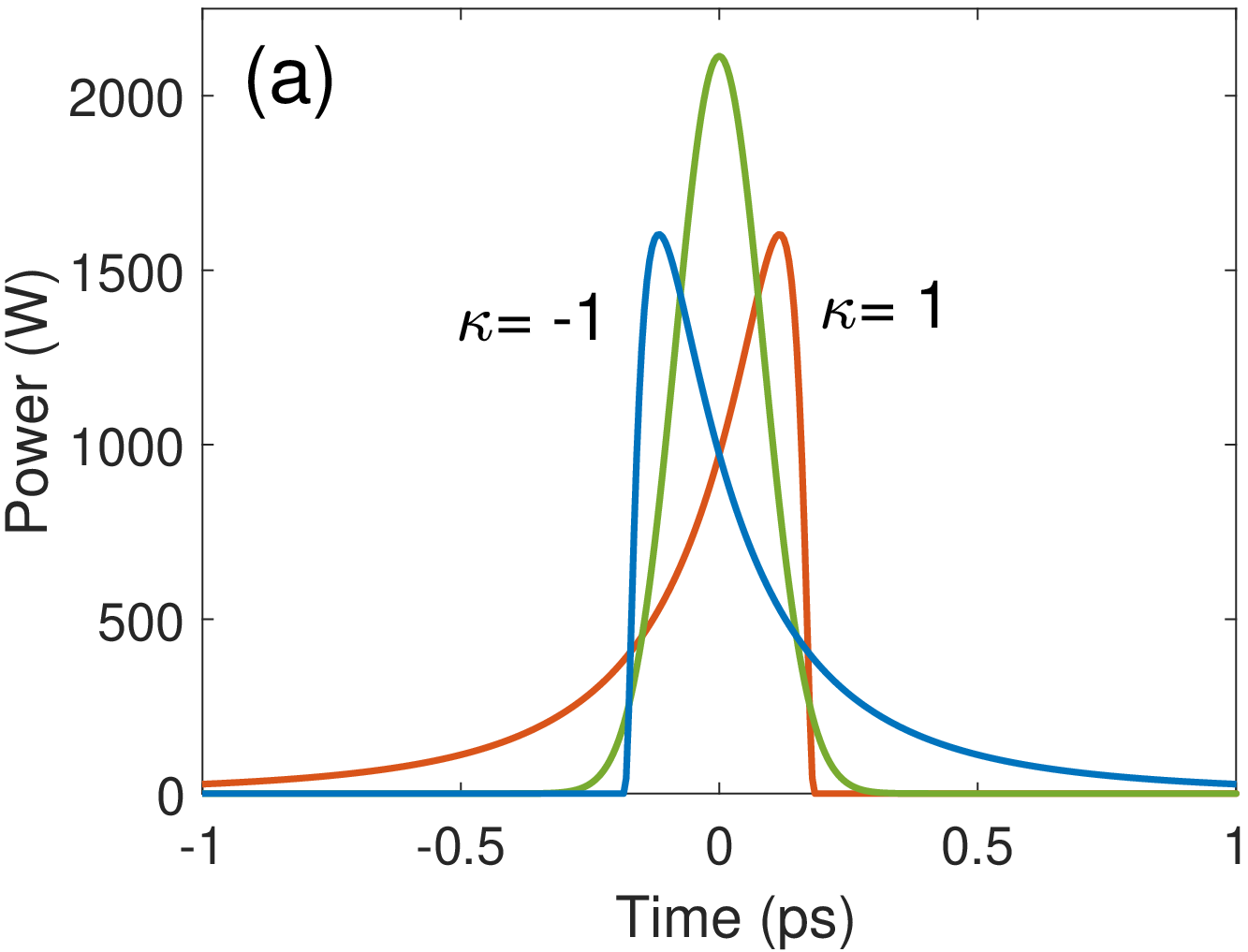}
\includegraphics[width=0.5\linewidth]{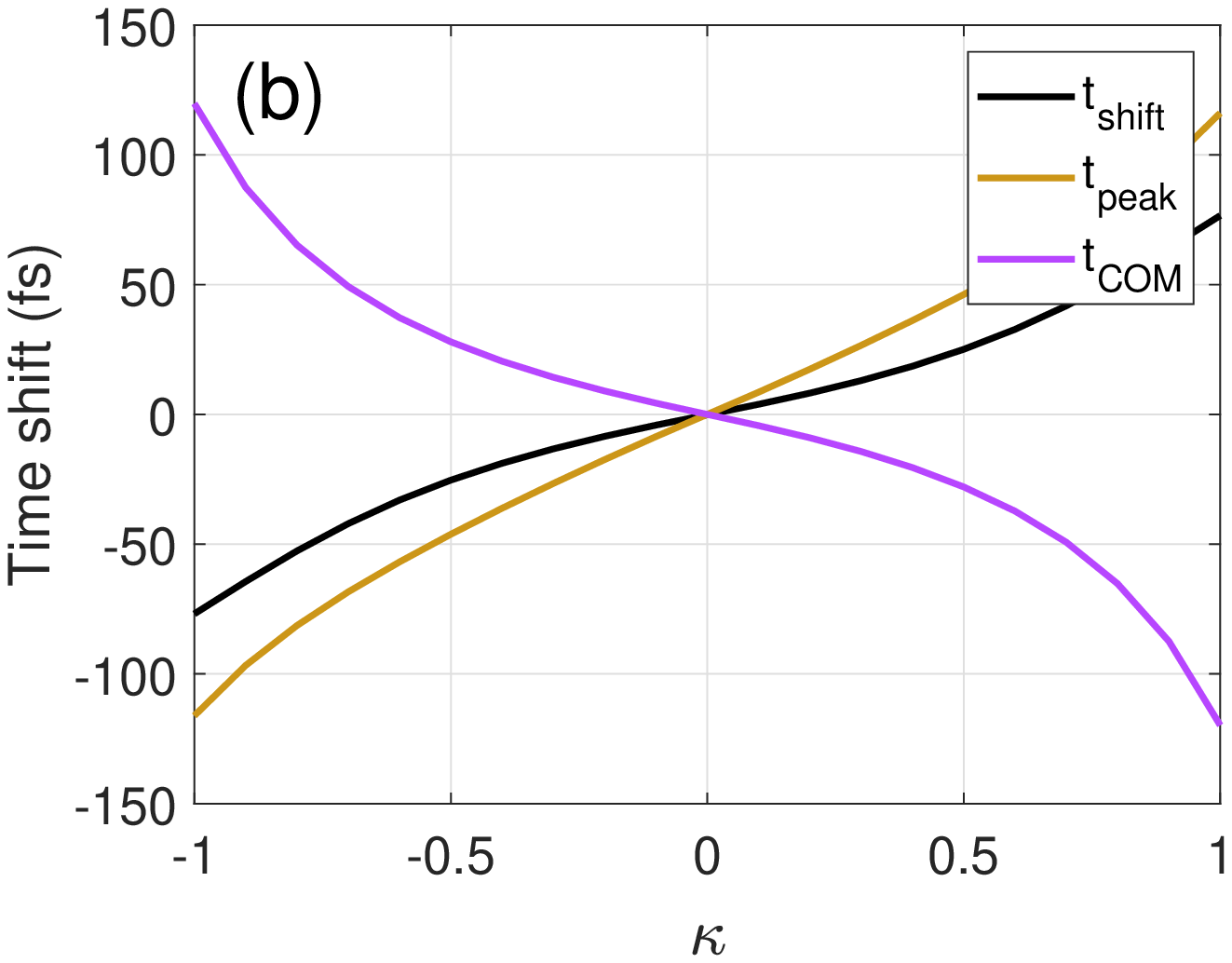}
\caption{Timing dependence on pulse shape. (a) The skew parameter $\kappa$ changes the shape of the generalized gaussian pulses. (b) Time shift of the signal ($t_{\textrm{shift}}$), pulse peak position ($t_{\textrm{peak}}$) and pulse 'center of mass' ($t_{\textrm{CoM}}$).}
\label{fig:skew}
\end{figure}

When reference timing pulses are used as depicted in Fig.~\ref{fig:schematic} the temporal drift of the stretched supercontinuum is subtracted out, that is, active stabilization is not needed in a 'full' setup. Other positive properties include insensitivity to power and pulse duration fluctuations. A negative feature is the fact that since the pulses are not temporally resolved, the pulse shapes do affect the readings. If the temporal shape of the timing pulses can be fully characterized, then numerical modelling can, of course, be used to correct the absolute timing. Ultrafast pulses can have very complicated shapes, but nevertheless, to gain some insight, Fig.~\ref{fig:skew} shows the change in timing for generalized Gaussian \cite{HoskingBook} pulse shapes (FWHM 200~fs) as the skew parameter $\kappa$ is varied. In this case, the time shift ($t_{\textrm{shift}}$) is roughly half the pulse peak position shift ($t_{\textrm{peak}}$) for $\vert\kappa\vert<0.5$ while the pulse 'center of mass' ($t_{\textrm{CoM}}$) moves in the opposite direction. Asymmetry in the pulse does lead to an asymmetry also in the XPM signal, but the effect is rather small and using it requires a better signal-to-noise ratio (SNR) than what was available in the present setup.


\section{Discussion}

The results above indicate that encouraging results can be obtained with a simple setup that is fully compatible with modern optical clockwork with only small additions. A simple approach to reduce the short term jitter is not to simply locate the peak of the pulse but to fit a suitable model (e.g. Eq. \ref{Eq:signal}) to the data, and the long term drift should be removed automatically in the full system of Fig. \ref{fig:schematic} due to the reference pulses.


For optimum performance, several shortcomings of the current proof-of-concept setup should be addressed. Simple improvements include, e.g, a customized high-resolution  spectrometer with high-speed read out (>100~klines/s available) that would reduce the dead time and allow a more detailed analysis of the signal, and a dual-channel spectrometer would enable simultaneous measurement of reference spectra for increased SNR. 
 If read-out rate is not important, then a high-resolution, compact echelle spectrograph using an imaging sensor could be used \cite{Probst2017a}. A polarization maintaining laser would increase stability, and a higher pulse repetition rate and/or an increase of the temporal stretching would enable a full pulse period to be covered. The supercontinuum should also be optimized for broad bandwidth and high spectral uniformity \cite{Heidt2011a}. Lowering the necessary pulse peak powers for XPM by several orders of magnitude should be possible using highly nonlinear waveguides or epsilon-near-zero materials \cite{Alam2018a}. For timing non-repetitive events, triggering of the detector or gating the supercontinuum in order not to 'flood' the detector in between the events is needed. Finally, calibration of the delay-wavelength dependence is probably easiest with the help of timing pulses from another mode-locked laser operating at a slightly different repetition rate. This would cause the timing pulses to slowly drift across the measurement range enabling  accurate characterization of the delay-wavelength dependence.

In conclusion, a very simple yet powerful concept for all-optical event timing at the femtosecond level has been presented that is compatible with contemporary optical clockwork. Since time is encoded into an optical signal using optical timing pulses without resorting to high-speed electronics and electronic interpolation, very low drift over long times and temperature ranges is to be expected.
\begin{backmatter}
\bmsection{Funding} Academy of Finland (decisions 296476 and 328389).
\bmsection{Acknowledgments} The work is part of the Academy of Finland Flagship Programme, Photonics Research and Innovation (PREIN), decision 320168.
\bmsection{Disclosures} The authors declare no conflicts of interest.
\bmsection{Data availability} Data underlying the results presented in this paper may be obtained from the authors upon reasonable request.
\end{backmatter}

\bibliography{sample}

\begin{thebibliography}{10}
\newcommand{\enquote}[1]{``#1''}

\bibitem{Kalisz2003a}
J.~Kalisz, {\protect\JournalTitle{Metrologia}} \textbf{41}, 17 (2003).

\bibitem{Keranen2011a}
P.~Keranen, K.~Maatta, and J.~Kostamovaara, {\protect\JournalTitle{IEEE
  Transactions on Instrumentation and Measurement}} \textbf{60}, 3162 (2011).

\bibitem{Schiller2002a}
S.~Schiller, {\protect\JournalTitle{Opt. Lett.}} \textbf{27}, 766 (2002).

\bibitem{Deschenes2016a}
J.-D. Desch\^enes, L.~C. Sinclair, F.~R. Giorgetta, W.~C. Swann, E.~Baumann,
  H.~Bergeron, M.~Cermak, I.~Coddington, and N.~R. Newbury,
  {\protect\JournalTitle{Phys. Rev. X}} \textbf{6}, 021016 (2016).

\bibitem{Abuduweili2020a}
A.~Abuduweili, X.~Chen, Z.~Chen, F.~Meng, T.~Wu, H.~Guo, and Z.~Zhang,
  {\protect\JournalTitle{Opt. Express}} \textbf{28}, 39400 (2020).

\bibitem{Kolner1989a}
B.~H. Kolner and M.~Nazarathy, {\protect\JournalTitle{Opt. Lett.}} \textbf{14},
  630 (1989).

\bibitem{Foster2008a}
M.~A. Foster, R.~Salem, D.~F. Geraghty, A.~C. Turner-Foster, M.~Lipson, and
  A.~L. Gaeta, {\protect\JournalTitle{Nature}} \textbf{456}, 81 (2008).

\bibitem{Li2021a}
B.~Li, J.~Bartos, Y.~Xie, and S.-W. Huang, {\protect\JournalTitle{Optica}}
  \textbf{8}, 1109 (2021).

\bibitem{Chou2007a}
J.~Chou, O.~Boyraz, D.~Solli, and B.~Jalali, {\protect\JournalTitle{Applied
  Physics Letters}} \textbf{91}, 161105 (2007).

\bibitem{Galvanauskas1992a}
A.~Galvanauskas, J.~A. Tellefsen, A.~Krotkus, M.~Öberg, and B.~Broberg,
  {\protect\JournalTitle{Applied Physics Letters}} \textbf{60}, 145 (1992).

\bibitem{Valdmanis1986a}
J.~Valdmanis, \enquote{Real time picosecond optical oscilloscope,} in
  \emph{ultrafast Phenomena V,}  G.~Fleming and A.~Siegman, eds. (Springer, New
  York, 1986), p.~82.

\bibitem{Beddard1992a}
G.~Beddard, G.~McFadyen, G.~Reid, and J.~Thorne,
  {\protect\JournalTitle{Chemical Physics Letters}} \textbf{198}, 641 (1992).

\bibitem{Jiang1998b}
Z.~Jiang and X.-C. Zhang, {\protect\JournalTitle{Applied Physics Letters}}
  \textbf{72}, 1945 (1998).

\bibitem{Chien2000a}
C.~Y. Chien, B.~L. Fontaine, A.~Desparois, Z.~Jiang, T.~W. Johnston, J.~C.
  Kieffer, H.~P\'{e}pin, F.~Vidal, and H.~P. Mercure,
  {\protect\JournalTitle{Opt. Lett.}} \textbf{25}, 578 (2000).

\bibitem{Leblanc2000a}
S.~P.~L. Blanc, E.~W. Gaul, N.~H. Matlis, A.~Rundquist, and M.~C. Downer,
  {\protect\JournalTitle{Opt. Lett.}} \textbf{25}, 764 (2000).

\bibitem{Geindre2001a}
J.-P. Geindre, P.~Audebert, S.~Rebibo, and J.-C. Gauthier,
  {\protect\JournalTitle{Opt. Lett.}} \textbf{26}, 1612 (2001).

\bibitem{Kim2002a}
K.~Y. Kim, I.~Alexeev, and H.~M. Milchberg, {\protect\JournalTitle{Applied
  Physics Letters}} \textbf{81}, 4124 (2002).

\bibitem{Sharma2012a}
G.~Sharma, K.~Singh, I.~Al-Naib, R.~Morandotti, and T.~Ozaki,
  {\protect\JournalTitle{Opt. Lett.}} \textbf{37}, 4338 (2012).

\bibitem{Wahlstrand2016a}
J.~K. Wahlstrand, S.~Zahedpour, and H.~M. Milchberg, {\protect\JournalTitle{J.
  Opt. Soc. Am. B}} \textbf{33}, 1476 (2016).

\bibitem{AgrawalBook}
G.~Agrawal, \emph{Nonlinear Fiber Optics} (Academic Press, 2007).

\bibitem{Barnes1990a}
J.~Barnes and D.~Allan, {\protect\JournalTitle{NIST Technical Note 1318}}
  (1990).

\bibitem{Sullivan1990a}
D.~Sullivan, D.~Allan, D.~Howe, and F.~Walls, {\protect\JournalTitle{NIST
  Technical Note 1337}}  (1990).

\bibitem{HoskingBook}
J.~Hosking and J.~Wallis, \emph{Regional frequency analysis: an approach based
  on L-moments} (Cambridge University Press, 1997).

\bibitem{Probst2017a}
R.~A. Probst, T.~Steinmetz, Y.~Wu, F.~Grupp, T.~Udem, and R.~Holzwarth,
  {\protect\JournalTitle{Applied Physics B}} \textbf{123}, 76 (2017).

\bibitem{Heidt2011a}
A.~M. Heidt, A.~Hartung, G.~W. Bosman, P.~Krok, E.~G. Rohwer, H.~Schwoerer, and
  H.~Bartelt, {\protect\JournalTitle{Opt. Express}} \textbf{19}, 3775 (2011).

\bibitem{Alam2018a}
M.~Z. Alam, S.~A. Schulz, J.~Upham, I.~De~Leon, and R.~W. Boyd,
  {\protect\JournalTitle{Nature Photonics}} \textbf{12}, 79 (2018).

\end{thebibliography}


\end{document}